\documentclass[12pt]{iopart}
\usepackage[pdftex]{graphicx}
\usepackage{iopams}
\usepackage{cite}
\usepackage{hhline}

\renewcommand{\theequation}{\arabic{section}.\arabic{equation}}

 \def\1{\mathsf{1}}
 \def\({\left(}
 \def\){\right)}

\def\-{{\bf --}} 
\newcommand{\bra}[1]{\ensuremath{\langle #1 |}}
                                                                                
\newcommand{\ket}[1]{\ensuremath{| #1 \rangle}}
                                                                                
\newcommand{\braket}[2]{\ensuremath{\langle #1 | #2 \rangle}}
                                                                                
\newcommand{\braopket}[3]{\ensuremath{\langle #1 | #2 | #3 \rangle}}
                                                                                
\begin{document}

%%%%%%%%%%%%%%% TITLE AREA %%%%%%%%%%%%%%% 

\title{The PASEP at $q=-1$}
\date{July 2012}

\author{D.\ A.\ Johnston and M. S. Stringer}
\address{Department of Mathematics and the Maxwell Institute for Mathematical
Sciences,
Heriot-Watt University, Riccarton, Edinburgh EH14 4AS, Scotland}

\begin{abstract}
We investigate the partially asymmetric exclusion process (PASEP)
with open boundaries
when the reverse hopping rate of particles $q=-1$, using a representation of the PASEP algebra related to the al-Salam Chihara polynomials. When 
$q=-1$ the representation is two-dimensional, which allows for straightforward calculation of the normalization, current and density. We note that these quantities  behave in an {\it a priori} reasonable manner
in spite of the apparently unphysical value of $q$ as the input, $\alpha$, and output, $\beta$, rates are varied 
over the physical range of $0$ to $1$. 

As is well known, another two dimensional representation
exists   when $0 < q < 1$  and $abq=1$,
where $a=(1-q)/\beta -1$ and $b=(1-q)/\beta-1$, 
and we compare the behaviour at $q=-1$ with this. 
An extension to generalized boundary conditions where particles may enter and exit at both ends is briefly outlined.
We also note that 
a different representation related to the q-harmonic oscillator  
does not admit a straightforward truncation when $q=-1$ and discuss
why this is the case from the perspective of a lattice path 
interpretation of the PASEP normalization.

\end{abstract}

\pacs{05.40.-a, 05.70.Fh, 02.50.Ey}

\maketitle

%%%%%%%%%%%%%%% MAIN TEXT %%%%%%%%%%%%%%% 

\section{Introduction}

Analytically continuing physical parameters such as temperature and field to unphysical, even complex values, is often done in the study of equilibrium statistical mechanical models, for instance in determining the Lee-Yang \cite{LYM,LY} and Fisher  \cite{LP,Fish} zeroes for partition functions whose scaling properties provide an alternative approach to characterizing phase transitions. It has been less commonly practised for non-equilibrium models, though Blythe and Evans have considered ``normalization zeroes'' for the asymmetric exclusion process (ASEP) by taking complex injection or removal rates at the boundary and recovered the known phase diagram for the non-equilibrium steady states \cite{BEZeroes1,BEZeroes2} from the scaling properties of these zeroes. More recently, Sasamoto and Williams \cite{Sas_Will} showed that choosing a negative injection rate at one boundary could relate the finite and semi-infinite
partially asymmetric exclusion process  (PASEP) over the region in which the finite PASEP was defined. In \cite{Sas_Will} the interesting general question of whether 
one can still give a probabilistic or physical meaning to the corresponding
stationary distribution if one makes one or more parameters negative or complex in  finite Markov chain models, such as the PASEP, was posed. 

The PASEP with a reverse hopping rate (defined below) of $q=-1$  appears to be an example of a model with a probabilistic interpretation since we find that not only does the normalization remain positive but quantities such as the current and particle density  continue to behave in a reasonable manner. An additional  motivation for choosing  $q=-1$ comes from consideration  of finite dimensional representations of the PASEP algebra. Such $n$-dimensional representations have been obtained when $0<q<1$ along  lines  in the non-equilibrium steady state phase diagram of the PASEP given by 
$a b q^{n-1} =1$, where $a = ( 1 -q )/\alpha -1$ and $b =(1-q)/\beta -1$ are convenient parametrizations of the injection and extraction rates $\alpha, \beta$. However, they also exist when $q^n=1$.  These representations have not been considered  before in the context of the PASEP precisely because the hopping rate takes an unphysical value but the study of the $XXZ$ spin chain, which is closely related to the PASEP, at $q^n=1$ has been a topic of interest for many years \cite{Nepo0,Nepo1,Nepo2,Doikou}.
The simplest non-trivial example of such finite representations in the PASEP is found when $q=-1$, giving the two dimensional representation whose properties we investigate here.

We adopt a low-tech approach throughout, diagonalizing the $2 \times 2$ matrices
of the PASEP algebra and explicitly calculating the current and density from these, which allows straightforward direct comparisons between the 
two-dimensional representations for $abq=1$ and $q=-1$. We briefly outline
the existence of similar finite dimensional representations for $q^n=1$ (and $q=-1$ in particular) when the 
boundary conditions are generalized. 
We also discuss the lattice path interpretation
of the generating function for the PASEP normalization  and note that a different representation of the PASEP algebra related to the q-harmonic oscillator does {\it not} admit  finite dimensional representations, at least in an obvious manner, because it is no longer possible to implement the boundary conditions.  

\section{Model Definition and Basic Properties}
We start with the standard setup for the PASEP with continuous time Markov dynamics on a finite lattice of length $N$ with open boundaries.
\begin{figure}[h]
\begin{center}
\includegraphics[scale=0.9]{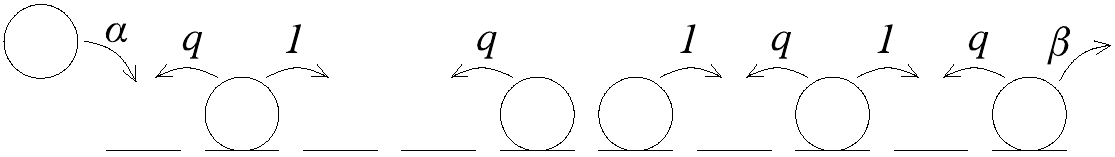}
\end{center}
\caption{\label{fig:PASEPrules}A typical particle configuration with
allowed moves and rates in the PASEP model. Particles are injected on the left at a rate $\alpha$ when a space is available and leave at a rate $\beta$ on the right. Their internal hopping rates are $1$ to the right and $q$ to the left when spaces are available.}
\end{figure}
Specifying the various injection, extraction and hopping rates for particles as shown in Fig.\,(1) defines the model. The model is clearly out of equilibrium since it supports a current driven by the injection and extraction of particles but
non-equilibrium steady states  can exist.
The matrix product 
ansatz solution for the PASEP, \cite{DEHP,PASEP,Blythe_review} calculates the probability $P(\mathcal{C})$ of a given 
configuration $\mathcal{C}$  in a non-equilibrium steady state  by employing
two vectors $\bra{W}$ and $\ket{V}$ to sandwich a word
$X_1 X_2 \ldots X_N$ generated from an alphabet of two matrices $D,E$ representing 
particles and holes in the configuration $\mathcal{C}$. This gives
\begin{equation}
\label{eqn:Pansatz}
P(\mathcal{C})=\frac{\braopket{W}{X_1 X_2 \ldots X_N}{V}}{Z_N} \;.
\end{equation}
where the normalization for lattice of length $N$ is
\begin{equation}
\label{eqn:Zdef}
Z_N=\braopket{W}{(D+E)^N}{V}=\braopket{W}{C^N}{V} \;,
\end{equation}
and we have defined $C=D+E$. These expressions are then related to the dynamics 
of the PASEP  by imposing a quadratic algebra and boundary conditions on the matrices 
$D,E$ and vectors $\bra{W}$,$\ket{V}$
\begin{eqnarray}
\label{eqn:DEcommute}
DE-qED &=& D+E \;, \nonumber \\
\label{eqn:EonW}
\alpha \bra{W}E &=& \bra{W} \;,\\
\label{eqn:DonV}
\beta D \ket{V} &=& \ket{V}  \; . \nonumber
\end{eqnarray}
Exact solutions for any $N$ may then be obtained by finding representations
of the matrices and vectors or by normal-ordering the matrices to let them
act on the appropriate boundary vector. The end result \cite{PASEP} is a phase diagram for the non-equilibrium steady states  of the PASEP which contains a low-density ($LD$)
a high density ($HD$) and a maximal current ($MC$) phase as the boundary rates $\alpha,\beta$
are varied, with the transition lines between the various phases being shown in Fig.\,(2). A novel feature is that a one-dimensional non-equilibrium system
can exhibit (boundary driven) phase transitions, unlike its equilibrium counterparts in one dimension such as the Ising model. 
\begin{figure}[h]
\begin{center}
\includegraphics[scale=0.35]{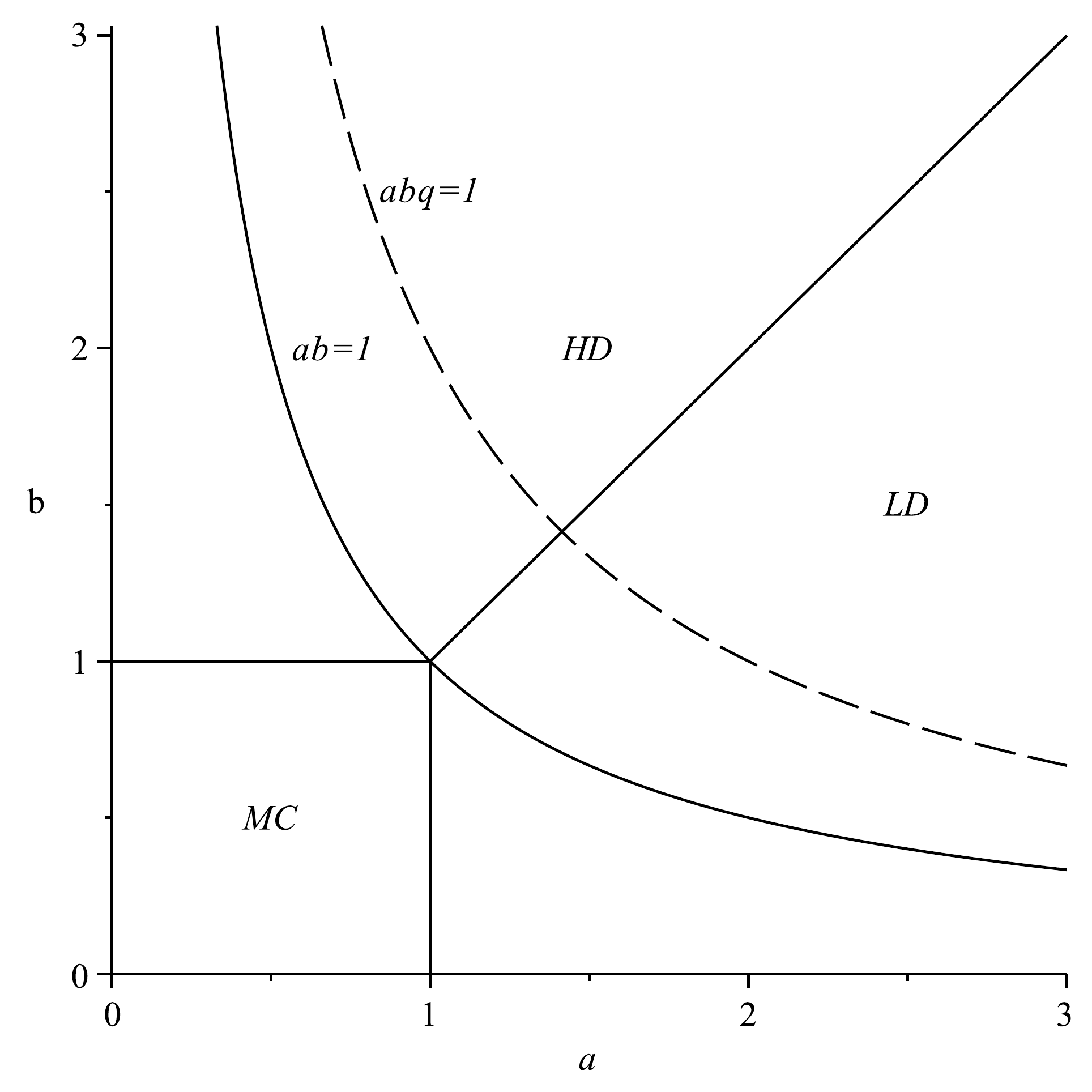}
\caption{\label{basicpd} The phase diagram of the PASEP. Here HD, LD and MC denote the high-density, low-density and maximal-current phases, respectively. The mean field line (a hyperbola) $ab=1$ is indicated, below which no finite dimensional representations exist when $0<q<1$. We also show
the dashed line $abq=1$ along which the physical two dimensional representation may be defined,  taking $q=1/2$ for definiteness.}
\end{center}
\end{figure}

In general, representations of the PASEP algebra are non-unique and will be infinite dimensional. The particular choice of representation which we find useful here is related to al-Salam Chihara  \cite{alSal} polynomials and is given by \cite{SAS,USW,SMW}
\begin{equation}
\bra{W} =  (1,0,0,\cdots ) \qquad
\ket{V} =  (1,0,0,\cdots )^T \;,
\end{equation}
and
\begin{eqnarray}
D &=& {1 \over 1 - q} \left(
\begin{array}{cccc}
1+ b    & \sqrt{c_1}    & 0             & \cdots\\
0     & 1 + b q  & \sqrt{c_2}    & {}\\
0               & 0     & 1 + b q^2  & \ddots\\
\vdots          & {}            & \ddots        & \ddots
\end{array}
\right) \;,
\nonumber \\
E &=& {1 \over 1 - q} \left(
\begin{array}{cccc}
1 + a    & 0    & 0             & \cdots\\
\sqrt{c_1}       & 1 + a q  &  0    & {}\\
0               & \sqrt{c_2}     & 1 + a q^2  & \ddots\\
\vdots          & {}            & \ddots        & \ddots
\end{array}
\right) \;.
\label{eqn:repde2}
\end{eqnarray}
where the various parameters which appear are defined as
\begin{eqnarray}
a &=& { 1 - q \over \alpha} -1 \, , \nonumber \\
b &=& { 1 - q \over \beta} -1 \, ,\\
c_n &=& (1 - q^n ) ( 1 - a b q^{n-1} ) \nonumber \, .
\end{eqnarray}
With these choices the matrix $C$  is 
\begin{equation}
\label{C}
C=D +E = {1 \over 1 - q} \left(
\begin{array}{cccc}
2+ a +b   & \sqrt{c_1}    & 0             & \cdots\\
 \sqrt{c_1}    & 2 + (a + b )q  & \sqrt{c_2}    & {}\\
0               &   \sqrt{c_2}   & 2 + (a + b) q^2  & \ddots\\
\vdots          & {}            & \ddots        & \ddots
\end{array}
\right) 
\end{equation}
and the normalization $Z_N$ in equ.\,(\ref{eqn:Zdef}) may then be calculated using various techniques. 
Similar expressions involving $C,D$ and $E$, which we discuss below explicitly for both $abq = 1$ and $q = -1$, may also be 
evaluated for the current and density.

In the sequel we first reiterate the known results for the physical two-dimensional representation which exists when $abq=1$, before going on to consider $q=-1$. We compare the features of physical quantities such as the current and densities in both these representations and also discuss
the representation related to the q-harmonic oscillator where, at least naively, continuation to $q=-1$ does not appear to be possible.
We conclude with some general comments on other representations
with $q^n=1$ for $n>2$ and various possible generalizations.

\section{The physical two-dimensional representation: $abq=1$, $0<q<1$}

The study of finite dimensional representations of the PASEP algebra and other quadratic algebras associated with different reaction-diffusion models pre-dated the exact  matrix product solution of the PASEP \cite{ER,MS,Jafar}. It is useful to think of $C$ as a transfer matrix (indeed, when the PASEP is mapped onto a lattice path model, as discussed in section 7, it {\it is} a transfer matrix)
so looking at equ.\,(\ref{C}) we can see that an $n$-dimensional representation will exist if $c_n = (1 - q^n ) ( 1 - a b q^{n-1} )=0$ since this decouples the matrix into blocks. Only the block in the upper left-hand corner will give  a non-zero scalar product with the vectors $\bra{W}, \ket{V}$.
One way this can occur is if $a b q^{n-1}=1$, which restricts us to hyperbolic lines in the $a,b$ plane of the PASEP phase diagram for physical values of $q$ (i.e. $0 < q < 1$). The simplest of these is $n=1$, giving $ab =1$
which is just the mean field line for the PASEP, as indicated on Fig.\,(2). All the other finite dimensional representations may be defined on hyperbolae which lie above this, for instance the two-dimensional representation when $abq=1$, which is shown in Fig.\,(2) for $q=1/2$.

It is instructive to compare the two-dimensional representation at $q=-1$ with this other, physical, two-dimensional representation which exists along the line $abq=1$ for $0  < q < 1$. This latter has been discussed by both Essler and Rittenberg
\cite{ER} and Mallick and Sandow \cite{MS} in some generality,  including the possibility
of extraction at a rate $\gamma$ at the left boundary and injection at a rate $\delta$ at the right boundary. The overall structure of the phase diagram is not altered by this embellishment (at least when $q<1$) so we 
largely stick to the case of $\gamma=\delta=0$ for simplicity.

The line  $abq=1$ can cross the transition between the $HD$ and $LD$ phases in Fig.\,(2)
as $\alpha,\beta$ are varied, so we must choose either $a= 1/bq$ or $b = 1/ aq$ depending on the dominant eigenvalue. Taking the former for definiteness
(i.e. the $HD$ phase),  $C$ is given by
\begin{eqnarray}
C &=& \left(
\begin{array}{cc}
{1  + b^2 q + 2 bq  \over bq  (1-q) }& {i \over \sqrt{q}}   \\
 {i \over \sqrt{q}}     &  {1  + b^2  + 2 b\over b (1-q)}
\end{array}
\right) 
\end{eqnarray}
and  $D$ and $E$ by
\begin{eqnarray}
D &=&  \left(
\begin{array}{cc}
{1 + b  \over  (1-q)}   &  {i \over \sqrt{q}}   \\
  0     &  {1 + bq  \over  (1-q)}
\end{array}
\right) \nonumber \\
E &=&  \left(
\begin{array}{cc}
{1+ bq \over (1-q) bq}   &  0   \\
{i  \over \sqrt{q}}      & {1 + b\over (1 -q)b}
\end{array}
\right) \; .
\end{eqnarray}
The vectors $\bra{W}$ and $\ket{V}$ are simply truncated to two
components
\begin{equation}
\bra{W} =  (1,0) \qquad
\ket{V} =  (1,0)^T \; .
\end{equation}
For the $LD$ phase we exchange $b \leftrightarrow a$, and it is straightforward to verify  that both the quadratic algebra relation and boundary conditions
of equ.\,(\ref{eqn:DEcommute}) are still satisfied in both phases.

The off-diagonal elements of $C$ in this representation  are imaginary but the
expressions contributing to physical quantities such as the current and densities
are  real and positive. 
The eigenvectors of $C$ are
\begin{eqnarray}
\lambda_1 &=& {1 \over 1-q} \left( {1 \over b } + b + 2    \right) \nonumber \\
\lambda_2 &=& {1 \over 1-q} \left(  {1 \over b q } + bq  + 2    \right)
\end{eqnarray}
and it may thus be diagonalized  using
\begin{eqnarray}
S &=& \left(
\begin{array}{cc}
-i b \sqrt{q}   & 
-{i \over b \sqrt{q}} \\
1      & 
1
\end{array}
\right) 
\end{eqnarray}
and
\begin{eqnarray}
S^{-1} &=& { 1 \over b^2 q -1}  \left(
\begin{array}{cc}
 i  b \sqrt{q}  & 
 -1
  \\
-i  b \sqrt{q}   & 
b^2 q
\end{array}
\right) \; .
\end{eqnarray}
This gives
\begin{equation}
\Lambda = S^{-1} C S =\left(
\begin{array}{cc}
 \lambda_1 & 0   \\
0       & \lambda_2
\end{array}
\right) 
\end{equation}
and we can also define $\tilde{D}$ and $\tilde{E}$,
which are  both lower diagonal in the new basis \cite{ER}
\begin{eqnarray}
\label{DEabq}
\tilde{D} &=&  S^{-1} D S = \left(
\begin{array}{cc}
{ 1 + b \over (1-q)}  &  0  \\
-b   &  {1+bq \over (1-q) } 
\end{array}
\right) \nonumber \\
\tilde{E} &=& S^{-1} E S =  \left(
\begin{array}{cc}
{ 1 + b  \over (1-q)b }  &  0 \\
b   &  { 1 + bq \over  ( 1 -q)bq} 
\end{array}
\right) 
\end{eqnarray}
which, as we shall see below, simplifies the expressions for densities (and correlators) somewhat by comparison with $q=-1$. The vectors $\bra{\tilde{W}}, \ket{\tilde{V}}$ in the diagonalized basis are given by
\begin{eqnarray}
\bra{\tilde{W}} &=&   \bra{W} S = 
\left( -i  \sqrt{q} b   ,   -{ i \over \sqrt{q} b} \right) \nonumber \\ 
\ket{\tilde{V}} &=&    S^{-1} \ket{V} =  m \, (  1 ,  -  1)^T 
\end{eqnarray}
where 
\begin{eqnarray}
m &=&  {i b  \sqrt{q} \over b^2 q - 1}
\end{eqnarray}
and the combinations contributing to physical quantities again contrive 
to be real and positive. 

The first quantity of interest  is the normalization, which is given by
\begin{equation}
Z_N = \bra{W} C^N \ket{V} = \bra{W} S S^{-1} C^N S S^{-1} \ket{V} =  \bra{\tilde{W}} \Lambda^N \ket{\tilde{V}}
\end{equation}
and may be evaluated 
by using the definitions above  to give
\begin{eqnarray}
\label{ZN2}
Z_N &=& { 1 \over b^2 q -1 }  \lambda_1^N \left[ b^2 q - \left( {\lambda_2 \over \lambda_1 } \right)^N \right] .
\end{eqnarray}
In writing it in this form  we have noted that $\lambda_1>\lambda_2$, since $b>1/\sqrt{q}$ 
along $abq=1$ in the $HD$ phase.
We can thus define a correlation length
\begin{equation}
{1 \over \xi} =  \ln \left({ \lambda_1 \over \lambda_2} \right)
\end{equation}
and write the normalization as
\begin{equation}
\label{ZNtilde2}
Z_N =  \lambda_1^N \left[ \rho - \sigma  \exp \left(- {N \over  \xi }\right)  \right] 
\end{equation}
where
\begin{eqnarray}
\rho &=& {b^2 q \over ( b^2 q -1)} \nonumber \\
\sigma &=& {1 \over ( b^2 q -1)}
\end{eqnarray}
which clearly disentangles the bulk asymptotic term from the finite length corrections. 
The current $J_N$ is defined in the usual way as
\begin{equation}
J_N = {Z_{N-1} \over Z_N} 
\end{equation}
so asymptotically it is
\begin{equation}
J = J_{N \to \infty} = {1 \over \lambda_1} = (1-q){ b \over ( 1+b)^2}
\end{equation}
and we recover an expression for the current in the $HD$ phase which agrees with the mean-field value.

The density may be calculated in a similar manner to give
\begin{eqnarray}
\label{ZNtauL}
Z_N \left< \tau_j \right> &=& \bra{W} C^{j-1} D C^{N-j} \ket{V} \nonumber \\
&=& \bra{\tilde{W}} \Lambda^{j-1} \tilde{D}  \Lambda^{N-j} \ket{\tilde{V}}  
\end{eqnarray}
which may be written as
\begin{equation}
\label{ZNtau2}
Z_N \left< \tau_j \right> =  \lambda_1^{N-1}  
\left[ \mu -    \omega \exp \left( -{ (N -1) \over \xi} \right)
- \zeta \exp \left( -{ (j -1) \over \xi} \right) \right]
\end{equation}
where
\begin{eqnarray}
\mu &=&  {b^2 q (1+b) \over (1-q) (b^2 q -1)} \nonumber \\
\omega &=&  { (1 + bq)  \over (b^2 q -1)(1-q)  }   \nonumber \\
\zeta &=&  { b \over ( b^2 q -1) }  \; .
\end{eqnarray}
We can see from equ.\,(\ref{ZNtau2}) that asymptotically in $N$ the density $\tau$
starts at the value
\begin{equation}
\tau = { \mu \over \rho  \lambda_1} - { \zeta \over \rho \lambda_1 } \exp (- 1 / \xi )
\end{equation}
at the left boundary and increases exponentially to  
\begin{equation}
{ \mu \over \rho  \lambda_1}= { b \over 1+b}
\end{equation}
in the bulk. 
We plot the density profile in the $HD$ phase directly
calculated from equ.\,(\ref{ZNtauL}) in Fig.\,(3), where we have taken $b=4,q=1/2,N=100$, which shows this behaviour clearly.
\begin{figure}[h]
\begin{center}
\includegraphics[scale=0.35]{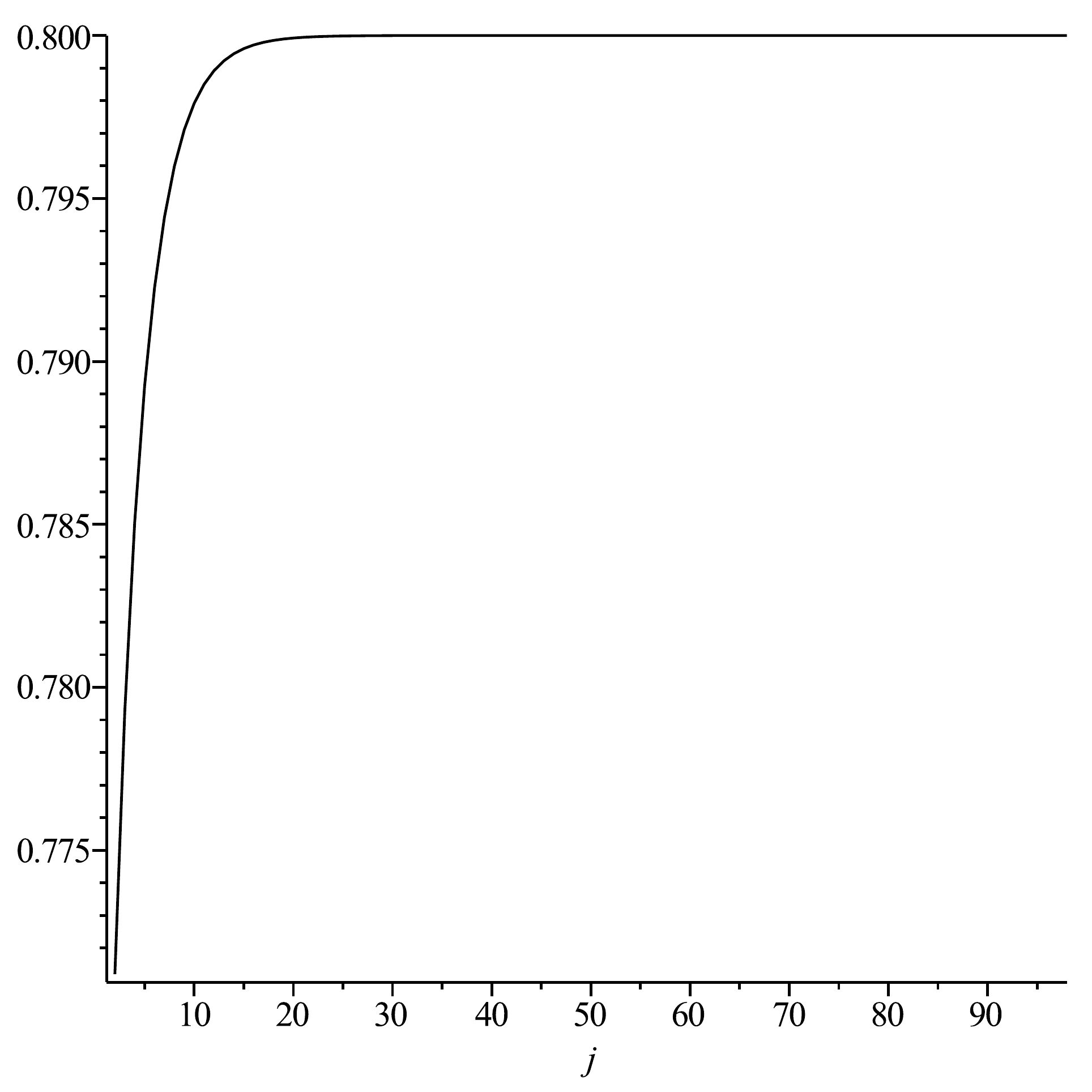}
\caption{\label{tauNabqj}The variation in the density $\left< \tau_j \right>$ along the lattice when $abq=1$,  for $b=4,q=1/2,N=100$ (in the $HD$ phase) showing the exponential increase to the bulk value of $4/5$ from the left hand boundary.}
\end{center}
\end{figure}
We may evaluate two point correlators in a similar manner, giving exponential correction terms which continue to play a role only near the left boundary for the connected correlator.

Similar expressions may be evaluated for the $LD$ phase \cite{ER}, here the density stays low in the bulk before increasing exponentially at the right boundary and the connected correlator corrections are non-zero only close to the right hand boundary. The results might be summarized by saying that for both the $HD$ and $LD$ phases one of the boundary rates ($b$ or $a$ respectively) dominates the behaviour and corrections to the bulk are seen at one of the boundaries 
(left or right respectively) only. The maximal current phase is inaccessible
to the two-dimensional representation along $abq=1$ as can be seen in Fig.\,(2).

\section{Another two-dimensional representation: $q=-1$}

Since $c_n=(1 - q^n ) ( 1 - a b q^{n-1} )=0$ when $q^n=1$, $n$-dimensional representations  also exist if $q$ is taken to be  a root of unity. Let us ignore any  qualms about the physical interpretation of such values of $q$ and
simply set $q=-1$ to find another two-dimensional representation with the following expressions for $C$
\begin{eqnarray}
C &=& \left(
\begin{array}{cc}
1+ {a +b\over 2}   & \frac{1}{2}\sqrt{2 - 2 a b}   \\
\frac{1}{2}\sqrt{2 - 2 a b}       & 1 - {a +b\over 2}
\end{array}
\right) 
\end{eqnarray}
and the individual $D,E$
\begin{eqnarray}
D &=& \left(
\begin{array}{cc}
\frac{1}{2}(1+ b)   &    \frac{1}{2}\sqrt{2 - 2 a b} \\
  0     & \frac{1}{2}(1 - b )
\end{array}
\right) \nonumber \\
E &=& \left(
\begin{array}{cc}
\frac{1}{2}(1+ a )   &  0   \\
\frac{1}{2}\sqrt{2 - 2 a b}       & \frac{1}{2}(1 - a )
\end{array}
\right) \; .
\end{eqnarray}
The expressions for the vectors remain the same as the other two-dimensional representation at $abq=1$
\begin{equation*}
\bra{W} =  (1,0) \qquad
\ket{V} =  (1,0)^T 
\end{equation*}
and both the quadratic algebra relation and boundary conditions
of equ.\,(\ref{eqn:DEcommute}) can be seen  to be  satisfied here too.

It is simpler to diagonalize $C$ for explicit calculations
just as it was for $abq=1$.
$C$ now has the eigenvectors
\begin{eqnarray}
\lambda_+ &=& 1 + \frac{1}{2}\sqrt{2 + a^2 + b^2} \nonumber \\
\lambda_- &=& 1 - \frac{1}{2}\sqrt{2 + a^2 + b^2}
\end{eqnarray}
so we may use
\begin{eqnarray}
S &=&  \left(
\begin{array}{cc}
{t \over r -a -b}   & -{t \over a+b+r}
  \\
1       & 
1
\end{array}
\right) 
\end{eqnarray}
and
\begin{eqnarray}
S^{-1} &=&  \left(
\begin{array}{cc}
 { t \over 2 r }  & 
 -{a + b -r \over 2 r}
  \\
-{ t \over 2 r}        &  { a + b +r\over 2 r}
\end{array}
\right) 
\end{eqnarray}
to diagonalize,
where we have defined $r=\sqrt{2 + a^2 + b^2}$ and $t= \sqrt{2-2ab}$ for conciseness. 
The natural range of the parameters $a,b$ when $q=-1$ is $1 \le a,b \le \infty$, since the
physical injection and extraction rates satisfy $0 \le \alpha,\beta \le 1$. The eigenvalues thus range over $2 \le \lambda_+ \le \infty$, $-\infty \le \lambda_- \le -1$.
We would therefore not expect any phase transitions, which require
$\lambda_+ = \lambda_-$, since this is not possible for physical values of $\alpha,\beta$ unlike the $abq=1$ representation. 

Diagonalizing, we have
\begin{equation}
\Lambda = S^{-1} C S =\left(
\begin{array}{cc}
 1 + \frac{r}{2} & 0   \\
0       & 1 - \frac{r}{2}
\end{array}
\right) 
\end{equation}
and similarly for $D,E$
\begin{eqnarray}
\label{DEminus1}
\tilde{D} &=& S^{-1} D S = \frac{1}{2} \left(
\begin{array}{cc}
1+ {1 + b^2 \over r }  &  -b + {1+b^2 \over r}  \\
-b - {1+b^2 \over r}      &  1-  {1 + b^2 \over r} 
\end{array}
\right) \nonumber \\
\tilde{E} &=&  S^{-1} E S = \frac{1}{2} \left(
\begin{array}{cc}
1+ {1 + a^2 \over r }  &  b - {1+b^2 \over r}  \\
b + {1+b^2 \over r}      &  1-  {1 + a^2 \over r} 
\end{array}
\right) \; .
\end{eqnarray}
The vectors $\bra{W}, \ket{V}$ in this basis are
\begin{eqnarray}
\bra{\tilde{W}} &=&   \bra{W} S = 
{1 \over t}( a + b +r  ,   a + b -r ) \nonumber \\ 
\ket{\tilde{V}} &=&    S^{-1} \ket{V} =  { t \over 2 r} (  1 ,  -  1)^T \; .
\end{eqnarray}
We can see that $\tilde{D}$ and $\tilde{E}$ in equ.\,(\ref{DEminus1}) are no longer lower diagonal, unlike the $\tilde{D}$ and $\tilde{E}$ in equ.\,(\ref{DEabq}) when $abq=1$, which will have consequences for the behaviour of the density and correlators. 

The first quantity of interest is again the normalization $Z_N$,
which may be evaluated to give
\begin{eqnarray}
\label{ZN}
Z_N &=& { 1 \over 2 r} ( \lambda_+)^N \left[ ( a + b + r )  - ( a + b - r) \left({ \lambda_- \over \lambda_+}\right)^N \right]  \; .
\end{eqnarray}
We can define the correlation length in this case by
\begin{equation}
{1 \over \tilde{\xi}} =  \ln \left({ \lambda_+  \over | \lambda_- |} \right)
\end{equation}
and write
\begin{equation}
\label{ZNrho}
Z_N =  \lambda_+^N \left[ \tilde{\rho} - (-1)^N \tilde{\sigma}  \exp \left(- {N \over  \tilde{\xi}}\right)  \right]
\end{equation}
with
\begin{eqnarray}
\tilde{\rho}  &=&   {a + b + r \over 2 r} = {a + b + \sqrt{2 + a^2 + b^2} \over 2 \sqrt{2 + a^2 + b^2}} \nonumber \\
\tilde{\sigma} &=&  {a + b - r \over 2 r} = {a + b - \sqrt{2 + a^2 + b^2} \over 2 \sqrt{2 + a^2 + b^2}} \; .
\end{eqnarray}
Although $\lambda_-$ is negative, $Z_N$ itself is positive for all $N$ (both even and odd) as can be seen by expanding equ.(\ref{ZN}). For instance, $Z_3$ is given by
\begin{eqnarray}
Z_3 &=& 5/2+(7/4)a+(7/4)b+(3/4)a^2+(3/4)b^2 \nonumber \\
&+& (1/8)a^3+(1/8)a^2 b+(1/8)a b^2+(1/8)b^3 \; .
\end{eqnarray}  
This in turn means that the current
\begin{equation*}
J_N = {Z_{N-1} \over Z_N}
\end{equation*}
is  a positive quantity. As we can see in Fig.\,(4)
\begin{figure}[h]
\begin{center}
\includegraphics[scale=0.35]{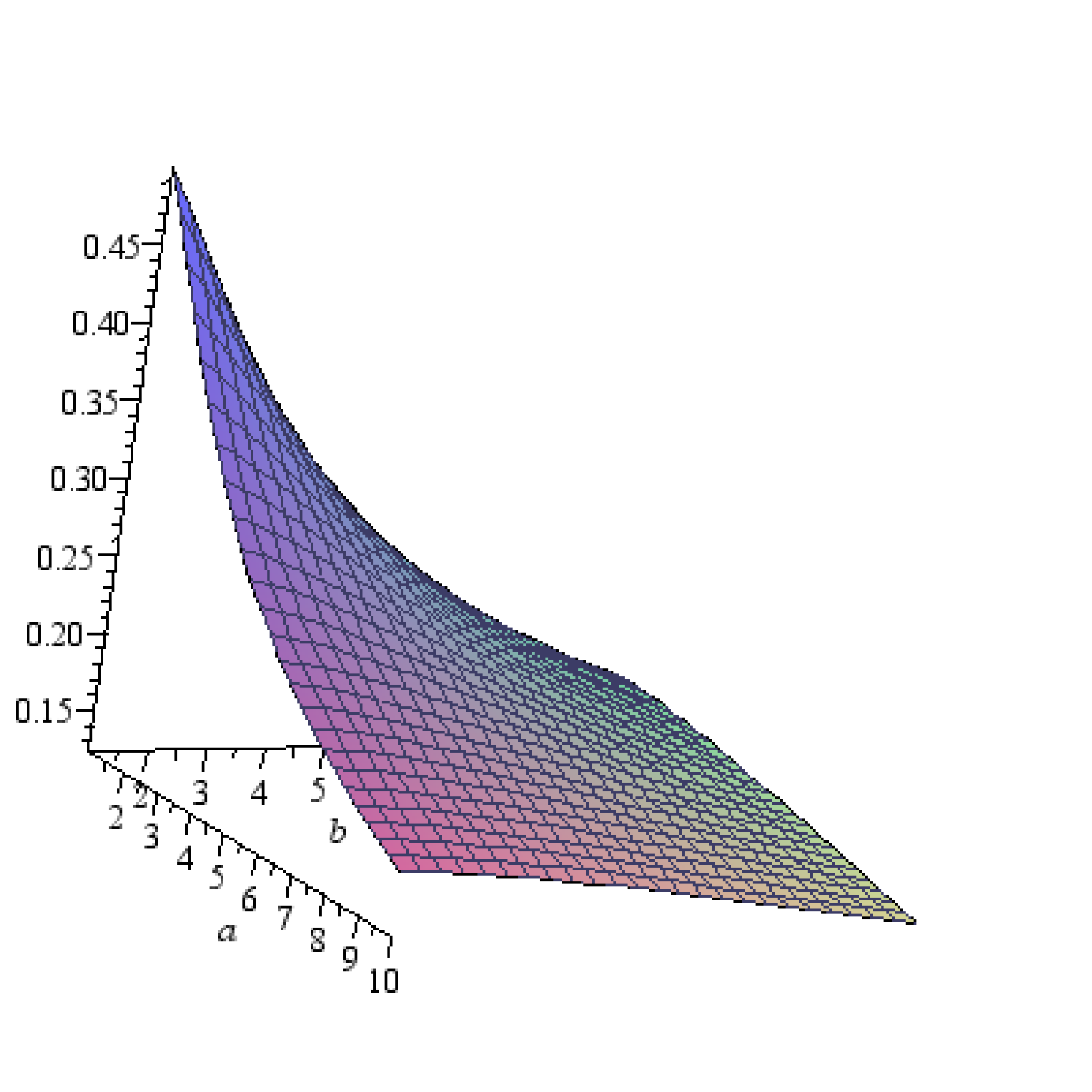}
\caption{\label{JN} The current $J_N$ for $N=100$, which is typical, plotted for $a,b=1 \ldots 10$}
\end{center}
\end{figure}
it is a smoothly decreasing function of $a,b$ dropping from a value of $J=1/2$
at $a=b=1$ (i.e. $\alpha=\beta=1$) to zero at large $a,b$. Asymptotically in $N$ it is given by
\begin{equation}
\label{Jasymp}
J = J_{N \to \infty} = {1 \over \lambda_+} = {1 \over 1 + \frac{1}{2}\sqrt{2 + a^2 + b^2} } \; .
\end{equation}
Remembering that $a=2/\alpha -1$, $b=2/\beta-1$ when $q=-1$, increasing $a,b$ from $a=b=1$  corresponds to decreasing the injection and extraction rates $\alpha,\beta$ so the observed falloff of $J$ and the values it takes in Fig.\,(4) are not physically unreasonable.

When $q=-1$ the current $J$  in equ.\,(\ref{Jasymp}) depends on both $a$ and $b$, which is not the case 
when $0 < q < 1$. For the latter  it is given by
$(1-q)a/(1+a)^2 $ and $(1-q)b/(1+b)^2$ in the low and high density phases respectively  and by $J = (1-q)/4$ in the maximal current phase. A naive continuation to $q=-1$ from the maximal current phase which extends to $\alpha=\beta=1$ gives the value of $J=1/2$
observed here for $q=-1$ at $\alpha=\beta=1$. $Z_N$ and hence $J_N$ contain exponential corrections which oscillate in sign, a feature which is also apparent in the density, which we turn to next.

If we evaluate the density
\begin{eqnarray}
\label{tauFig5}
\left< \tau_j \right> &=& {\bra{W} C^{j-1} D C^{N-j} \ket{V} \over Z_N} \nonumber \\
&=& {\bra{W} S^{-1} ( S C^{j-1} S^{-1} ) ( S D S^{-1} ) ( S C^{N-j} S^{-1} ) S \ket{V} \over Z_N } \nonumber \\
&=&  {\bra{\tilde{W}} \Lambda^{j-1} \tilde{D}  \Lambda^{N-j} \ket{\tilde{V}} \over Z_N} \; .
\end{eqnarray}
the individual terms in the numerator  no longer remain positive,
for example
\begin{eqnarray}
\bra{W} C D C \ket{V} &=& 5/4+b+(3/4) a+(5/8) b^2 \nonumber \\
&+& (1/8) a^2 \mathbf{-(1/8) a^2 b}+(1/4) a b^2+(1/8) b^3 
\end{eqnarray}
but the numerator as a whole {\it does} stay positive. In general
\small
\begin{eqnarray}
Z_N \left< {\tau_j} \right> &=& \bra{W} C^{j-1} D C^{N-j} \ket{V} \nonumber \\
&=& { 
(a+ b +r) \lambda_+^{N-1} \over 4r} \left[  \left( 1 + {1+b^2 \over r}\right) + \left( {\lambda_- \over  \lambda_+}\right)^{N-j} \left( b - { 1 +b^2 \over r} \right) \right]
 \nonumber \\
{} \\
&-& { (a+ b -r) \lambda_+^{N-1} \over 4r}  \left[  \left( 1 -{1+b^2 \over r}\right) \left({\lambda_- \over \lambda_+} \right)^{N-1} + \left({ \lambda_- \over \lambda_+} \right)^{j-1} \left( b + { 1 +b^2 \over r} \right) \right] 
\nonumber
\end{eqnarray}
\normalsize
which, if we adopt a similar notation to equ.\,(\ref{ZNrho}),
may be written as
\begin{eqnarray}
\label{ZNtau}
Z_N \left< {\tau_j} \right> &=&  \lambda_+^{N-1}  
\left[ \tilde{\mu} + (-1)^{N-j} \tilde{\nu} \exp \left(-{(N-j) \over \tilde{\xi}} \right) \right. \nonumber \\
{}\\
&-& \left.  \tilde{\omega} (-1)^{N-1} \exp \left( -{ (N -1) \over \tilde{\xi}} \right)
- (-1)^{j-1} \tilde{\zeta} \exp \left( -{ (j -1) \over \tilde{\xi}} \right)
\right] \; .
\nonumber
\end{eqnarray}
This should be contrasted with the corresponding expression for $abq=1$ in equ.\,(\ref{ZNtau2}), which contains one less exponential term because of the
lower diagonal form of $\tilde{D}$ in that case. 
The various coefficients in the above are given by
\begin{eqnarray}
\tilde{\mu} &=& {(a+ b + \sqrt{2 + a^2 + b^2})  \over 4 \sqrt{2 + a^2 + b^2} }   
\left( 1 + {1+b^2 \over \sqrt{2 + a^2 + b^2}} \right)  \nonumber \\
\tilde{\nu} &=&  {(a+ b + \sqrt{2 + a^2 + b^2})  \over 4 \sqrt{2 + a^2 + b^2} }   
\left( b - {1 + b^2 \over \sqrt{2 + a^2 + b^2}} \right)  \nonumber \\
\tilde{\omega} &=& {(a+ b - \sqrt{2 + a^2 + b^2})  \over 4 \sqrt{2 + a^2 + b^2} }   
\left( 1 - {1+b^2 \over \sqrt{2 + a^2 + b^2}} \right) \\
\tilde{\zeta} &=&  
{(a+ b - \sqrt{2 + a^2 + b^2})  \over 4 \sqrt{2 + a^2 + b^2} }   
\left( b + {1+b^2 \over \sqrt{2 + a^2 + b^2}} \right) \; . \nonumber 
\end{eqnarray}
As with the current, the behaviour of the central (bulk) density appears to be physically reasonable.
From equ.\,(\ref{ZNtau}) and Fig.\,(5) we can see that the central density $\left< \tau_{50} \right>$ depends on both $a$ and $b$, rather than just $b$ as in the $HD$ phase 
and $a$ in the $LD$ phase when $0 < q <1$. 
From Fig.\,(5) it is also apparent that $\left< \tau_{50} \right>$ decreases monotonically with increasing $a$ (i.e. decreasing injection rate) and increases monotonically with increasing $b$ (decreasing extraction rate).
\begin{figure}[h]
\begin{center}
\includegraphics[scale=0.35]{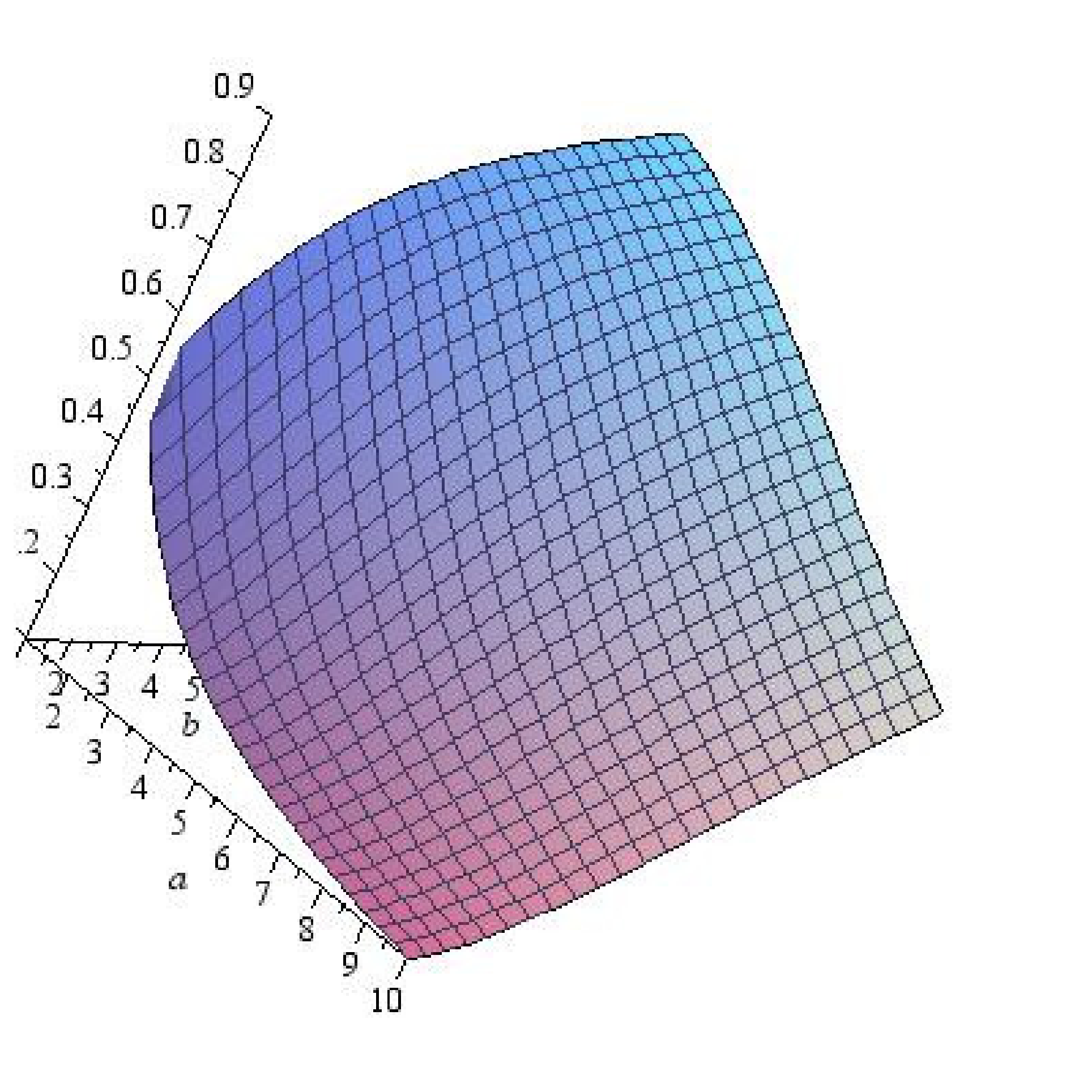}
\caption{\label{tauNab50} The central value of  $\left< \tau_j \right>$,  for $j=50,N=101$, calculated directly from equ.\,(\ref{tauFig5}) and plotted for $a,b=1 \ldots 10$.}
\end{center}
\end{figure}
The bulk value of the density is given asymptotically by
\begin{equation}
\label{taucentral}
\tau = { \tilde \mu \over \tilde \rho \lambda_+} = { 1 + b^2 + \sqrt{2 + a^2 + b^2} \over \sqrt{2 + a^2 + b^2} \, ( 2 + \sqrt{2 + a^2 + b^2})}
\end{equation}
so, in particular, we can see that $\tau=1/2$ when $a=b$, as evidenced by the central line along the surface in Fig.\,(5) and in the interior values in Fig\,(6). 
The asymptotic formula of equ.(\ref{taucentral}) is already an extremely good match to the directly calculated values shown in Fig.\,(5) when $N=101$ over the full range of $a,b$.
Note that, as is clear from Fig\,(5) and equ.\,(\ref{taucentral}), $\tau$  is not symmetric in $a,b$, unlike the current, but  the density of holes $1 - \tau$ is given by exchanging
$a \leftrightarrow b$
\begin{equation}
1- \tau  = { 1 + a^2 + \sqrt{2 + a^2 + b^2} \over \sqrt{2 + a^2 + b^2} \, ( 2 + \sqrt{2 + a^2 + b^2})} \; .
\end{equation}

If we now look at the density profile along the lattice for given $a,b$ there are oscillating positive and negative exponential corrections at {\it both} ends, when $j \sim 1$ and $j \sim N$, in contrast to the two-dimensional representation for $abq=1$ \cite{ER,MS} where corrections to the bulk are apparent at only one boundary.
\begin{figure}[h]
\begin{center}
\includegraphics[scale=0.35]{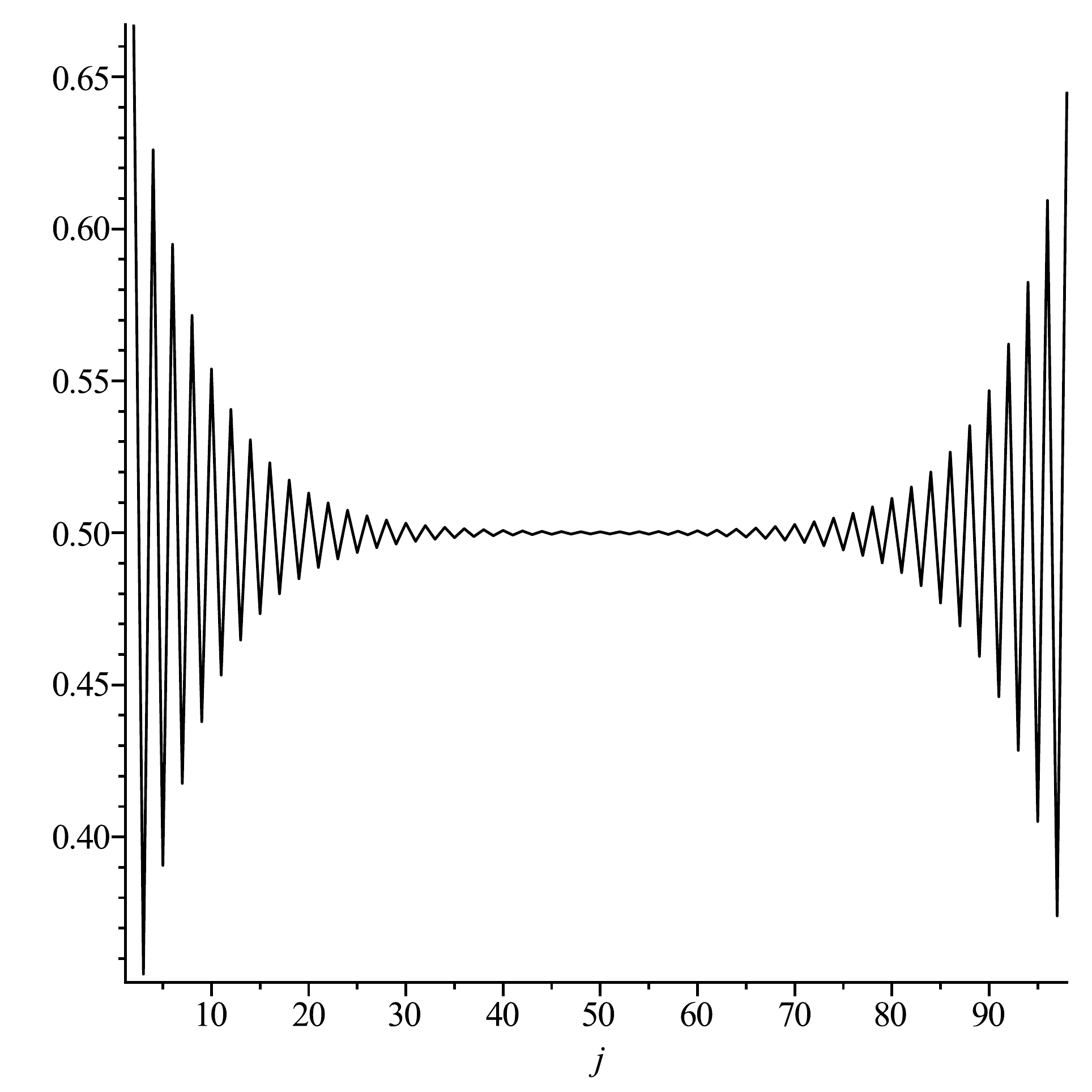}
\caption{\label{tauNabj}The variation in the density $\left< \tau_j \right>$ along the lattice when $q=-1$,  for $a=20,b=20,N=100$, showing the oscillating sign exponential corrections at both ends. The large $a,b$ values were chosen to make the oscillating corrections clearly visible.}
\end{center}
\end{figure}
The oscillating signs  when $q=-1$ in the correction terms  to the bulk density will still be present even if we consider only odd or even length lattices. A similar story holds for two point correlators, which continue to display ``two-sided'' oscillating sign corrections, in contrast to the corrections in the $HD$ and $LD$ phases when $abq=1$ discussed in the previous section.

\section{Finite dimensional representations when $\gamma, \delta \ne 0$} 

The PASEP boundary conditions may be generalized to allow 
wrong direction injection and removal rates.
Particles are now injected at a rate  $\alpha$ and removed 
at a rate $\gamma$
at the left boundary and particles
are removed at a rate $\beta$ and injected at a rate $\delta$ 
at the right boundary.  
Tridiagonal representations of $D$ and $E$ with 
non-zero $\gamma$ and $\delta$ still exist  \cite{USW} and they
may be written as
\begin{eqnarray}
D &=& {1 \over 1 - q} \left(
\begin{array}{cccc}
1 + d_0^{\natural}    & d_0^{\sharp}    & 0             & \cdots\\
d_0^{\flat}     & 1 + d_1^{\natural}     & d_1^{\sharp}    & {}\\
0               & d_1^{\flat}     & 1 + d_2^{\natural}   & \ddots\\
\vdots          & {}            & \ddots        & \ddots
\end{array}
\right) \;,
\nonumber \\
E &=& { 1 \over 1 - q } \left(
\begin{array}{cccc}
1 + e_0^{\natural}    & e_0^{\sharp}    & 0             & \cdots\\
e_0^{\flat}       & 1 + e_1^{\natural}   &  e_1^{\sharp}    & {}\\
0               & e_1^{\flat}     & 1 + e_2^{\natural}   & \ddots\\
\vdots          & {}            & \ddots        & \ddots
\end{array}
\right) \;,
\label{eqn:repde2a}
\end{eqnarray}
\begin{eqnarray}
\bra{\tilde W} =  (1,0,0,\cdots ) \;, \qquad
\ket{\tilde V} =  (1,0,0,\cdots )^T \;,
\label{eqn:repWV2a}
\end{eqnarray}
where  $d_n^{\sharp}$, $d_n^{\natural}$, $e_n^{\natural}$, $e_n^{\flat}$  are given by
\small
\begin{eqnarray*}
d_n^\natural &=&
\frac{q^{n-1}}{(1-q^{2n-2}abcd)(1-q^{2n}abcd)} \\
&&\times[
bd(a+c)+(b+d)q-abcd(b+d)q^{n-1}-\{ bd(a+c)+abcd(b+d)\} q^n \\
&&-bd(a+c)q^{n+1}+ab^2 cd^2(a+c) q^{2n-1}+abcd(b+d)q^{2n} ] \;, \\
e_n^\natural &=&
\frac{q^{n-1}}{(1-q^{2n-2}abcd)(1-q^{2n}abcd)} \\
&&\times[
ac(b+d)+(a+c)q-abcd(a+c)q^{n-1}-\{ ac(b+d)+abcd(a+c)\} q^n \\
&&-ac(b+d)q^{n+1}+a^2 bc^2 d(b+d) q^{2n-1}+abcd(a+c)q^{2n} ] \;, 
\end{eqnarray*}
\normalsize
\begin{eqnarray}
&&d_n^\sharp =
\frac{1}{1-q^nac}\mathcal{A}_n \;,\qquad
e_n^\sharp =
-\frac{q^nac}{1-q^nac}\mathcal{A}_n \;,   \\
&&d_n^\flat =
-\frac{q^nbd}{1-q^nbd}\mathcal{A}_n \;,\qquad
e_n^\flat =
\frac{1}{1-q^nbd}\mathcal{A}_n \;, \nonumber 
\end{eqnarray}
which involve the further parameters
\begin{eqnarray}
a &=& \frac{1}{2 \alpha} \left[ ( 1 - q - \alpha + \gamma) + \sqrt{ ( 1 - q - \alpha + \gamma)^2 + 4 \alpha \gamma} \right] \;, \nonumber  \\
b &=& \frac{1}{2 \beta} \left[ ( 1 - q - \beta + \delta) + \sqrt{ ( 1 - q - \beta + \delta)^2 + 4 \beta \delta} \right] \;,  \nonumber \\
c &=& \frac{1}{2 \alpha} \left[ ( 1 - q - \alpha + \gamma) - \sqrt{ ( 1 - q - \alpha + \gamma)^2 + 4 \alpha \gamma} \right] \;,  \\
d &=& \frac{1}{2 \beta} \left[ ( 1 - q - \beta + \delta) - \sqrt{ ( 1 - q - \beta + \delta)^2 + 4 \beta \delta} \right] \; , \nonumber  \\
\mathcal{A}_n &=&
%\begin{array}{l}
\textstyle
\left[
\frac{(1-q^{n-1}abcd)(1-q^{n+1})
(1-q^nab)(1-q^nac)(1-q^nad)(1-q^nbc)(1-q^nbd)(1-q^ncd)}
{(1-q^{2n-1}abcd)(1-q^{2n}abcd)^2(1-q^{2n+1}abcd)}
\right]^{1/2} \;. \nonumber
%\end{array}
\end{eqnarray}
In this case the polynomials associated with the representation are the Askey-Wilson polynomials \cite{AW}, of which the al-Salam Chihara polynomials are a specialization.

We note that when $\gamma=\delta=0$, $a$ and $b$ revert to the definitions of previous sections and $c=d=0$.
Although the expressions above are considerably more complicated than the
$\gamma=\delta=0$ case the structure of $C,D$ and $E$ is still similar. The
diagonal and off-diagonal elements of $C$ are now given by
$\hat d_n$ and $\sqrt{\hat c_n}$ where
\begin{eqnarray}
\hat d_n = {2 + d_n^{\natural}+ e_n^{\natural} \over 1 -q } \;, \nonumber \\
\hat c_{n+1} =  {(d_n^{\sharp} + e_n^{\sharp}) 
( d_n^{\flat} + e_n^{\flat})  
\over ( 1 - q)^2 } \;.
\end{eqnarray}
Explicitly,
\small
\begin{eqnarray}
{} \nonumber \\
\hspace{-2.5cm}
\hat{c}_{n+1} =
\frac{(1-q^{n-1}abcd)(1-q^{n+1})
(1-q^nab)(1-q^nac)(1-q^nad)(1-q^nbc)(1-q^nbd)(1-q^ncd)}
{(1-q^{2n-1}abcd)(1-q^{2n}abcd)^2(1-q^{2n+1}abcd) (1 - q)^2} \nonumber \\
{}
\end{eqnarray}
\normalsize
and $n$-dimensional representations will still exist if one of the terms in the numerator of $ \hat{c}_{n}$ is zero. This may occur in various ways: $abq^{n-1}=1$ as before; variations thereof involving at least one of $c,d$ such as $acq^{n-1}=1$ or $cdq^{n-1}=1$; or even a four-parameter condition
$abcdq^{n-2}=1$. The latter two possibilities are not available when $\gamma=\delta=0$. 
We can also see that  $n$-dimensional representations will continue to exist when $q^n=1$, since a factor of $1-q^n$ remains  present in the numerator of $\hat c_n$.

Curiously, conditions  of the form $abcdq^{n}=1$ and $q^{n}=1$ are both encountered when solving the open PASEP using the Bethe ansatz \cite{Fab_bethe1, Fab_bethe2, Fab_bethe3, Simon1, Simon2}, but the exact relation between finite dimensional representations of the matrix ansatz (indeed, the matrix ansatz in general) and the Bethe ansatz solutions is unclear.  

\section{A representation that does not truncate gracefully at $q=-1$.}

An alternative representation the PASEP algebra
related to the q-harmonic oscillator \cite{Macf} and q-Hermite polynomials
was employed in the original solution of the PASEP in \cite{PASEP}.
In this $D$ and $E$ are written as
\begin{eqnarray}
D = { 1 \over 1-q} + { 1\over 1-q } \hat{a} \nonumber \\
E = { 1 \over 1-q} + { 1\over 1-q } \hat{a}^{\dagger} 
\end{eqnarray}
where $\hat{a}, \hat{a}^{\dagger}$ satisfy q-boson commutation relations as a consequence of the PASEP algebra
\begin{eqnarray}
\hat{a}\hat{a}^{\dagger} &-& \, \,  q  \hat{a}^{\dagger}  \hat{a} = 1 -q  \nonumber \\
\hat{a}^{\dagger} \ket{n} &=& \left( 1-q^{n+1}  \right)^{1/2} \ket{n+1} \nonumber \\
\hat{a} \ket{n} &=& \left( 1-q^n  \right)^{1/2} \ket{n-1}\nonumber \\
\hat{a} \ket{0} &=& \, \, 0 \; .
\end{eqnarray}
The matrices $D$ and $E$ are given explicitly in this representation by
\begin{eqnarray}
D &=& {1 \over 1 - q} \left(
\begin{array}{cccc}
1    & \sqrt{1- q}    & 0             & \cdots\\
0     & 1   & \sqrt{1 - q^2}    & {}\\
0               & 0     & 1   & \ddots\\
\vdots          & {}            & \ddots        & \ddots
\end{array}
\right) \;,
\nonumber \\
E &=& {1 \over 1 - q} \left(
\begin{array}{cccc}
1     & 0    & 0             & \cdots\\
\sqrt{1-q}       & 1   &  0    & {}\\
0               & \sqrt{1-q^2}     & 1   & \ddots\\
\vdots          & {}            & \ddots        & \ddots
\end{array}
\right) \;.
\label{eqn:repde2h}
\end{eqnarray}
and the vectors $\bra{W}, \ket{V}$ by
\begin{eqnarray}
\label{qvec}
\braket{W}{n} =   \kappa { a^n \over \sqrt{(q;q)_n}}  \nonumber \\
\braket{n}{V} =    \kappa { b^n \over \sqrt{(q;q)_n}} 
\end{eqnarray}
with
\begin{equation}
{1 \over \kappa^2} =   \sum_{n=0}^{\infty} { (ab)^n \over (q;q)_n} \; .
\end{equation}
We have used the standard notation for (shifted) $q$-factorials in the above
\begin{eqnarray}
\label{eqn:qfacdef}
(a;q)_n &=& \prod_{j=0}^{n-1} (1-aq^j) \; .
\end{eqnarray}
The boundary conditions
\begin{eqnarray}
\alpha \bra{W}E &=& \bra{W} \nonumber \\
\beta D \ket{V} &=& \ket{V} 
\end{eqnarray}
are implemented in a slightly different manner in this representation compared with the previous section. We can see that the action of $D$ on $\ket{V}$, for instance,   
\begin{eqnarray}
\label{Dq}
D \ket{V} &=& {\kappa \over 1 - q} \left(
\begin{array}{cccc}
1    & \sqrt{1- q}    & 0             & \cdots\\
0     & 1   & \sqrt{1 - q^2}    & {}\\
0               & 0     & 1   & \ddots\\
\vdots          & {}            & \ddots        & \ddots
\end{array}
\right)
\left( \begin{array}{c}
1 \\
{b \over \sqrt{1-q}} \\
{b^2 \over \sqrt{(1-q)(1-q^2)}}  \\
\vdots
\end{array}\right)
\end{eqnarray}
is to shift up a component of $\ket{V}$ and add it to the component above in order to satisfy
$ \beta D \ket{V} = \ket{V}$.

Although
the $q$-boson relations collapse to suitably scaled fermionic anti-commutation relations
for a two-state system
when $q=-1$
\begin{eqnarray}
\hat{a}\hat{a}^{\dagger} &+&  \hat{a}^{\dagger}  \hat{a} = 2 \nonumber \\
\hat{a}^{\dagger} \ket{0} &=&  \sqrt{2} \, \ket{1} \nonumber \\
\hat{a} \, \ket{1} &=& \sqrt{2}  \, \ket{0}\nonumber \\
\hat{a} \, \ket{0} &=& \, \,  \emptyset \nonumber \\
\hat{a}^{\dagger} \ket{1} &=& \, \, \emptyset
\end{eqnarray}
and the matrices $C,D$ and $E$ decouple into $2 \times 2$ blocks, the vectors, for instance $\ket{V}$ in equ.\,(\ref{Dq}), and normalization $\kappa$ can be seen to be ill-defined when $q=-1$. 
This particular representation 
is therefore not suitable for continuing to $q=-1$ (nor, indeed, $q=1$ \cite{Blythe_review}). In the next section, where the relation between
the PASEP and lattice path models is outlined, we see heuristically why this is so.

\section{The PASEP and Lattice Paths}
The generating function
of $Z_N$
\begin{equation}
{\cal Z}(z) = \sum_N Z_N z^N
\end{equation}
can be thought of as
a ``grand-canonical'' normalization 
where  the PASEP normalizations for various $N$ are combined with a fugacity $z$.
The closest singularities of ${\cal Z}(z)$ to the origin can then be used to determine
the asymptotic behaviour of $Z_N$, a common technique in analytic combinatorics
\cite{Wilf,Flaj}.

It is also useful to interpret ${\cal Z}(z)$
as the generating function for a model of weighted lattice paths \cite{Brak1,Brak2,Jafar0,Us1,Us2,Us3} ,
which gives some insight into the PASEP phase diagram. From this point of view
the tridiagonal matrix $C$ in the particular representation of equ.\,(\ref{C}) is a transfer matrix for Motzkin paths composed 
of diagonal up and down steps with with weights $\sqrt{c_n} z/ (1-q)$ at height $n$
and horizontal steps with weight $d_n z /(1-q)$ at height $n$. 
A path of length $n$ will be weighted by $z^n$ and some path-dependent product
of $c_n$ and $d_n$, so by tuning these via $\alpha$, $\beta$ and $q$ we can
change the dominant paths contributing to the ensemble.
\begin{figure}[h]
\begin{center}
\includegraphics[scale=0.45]{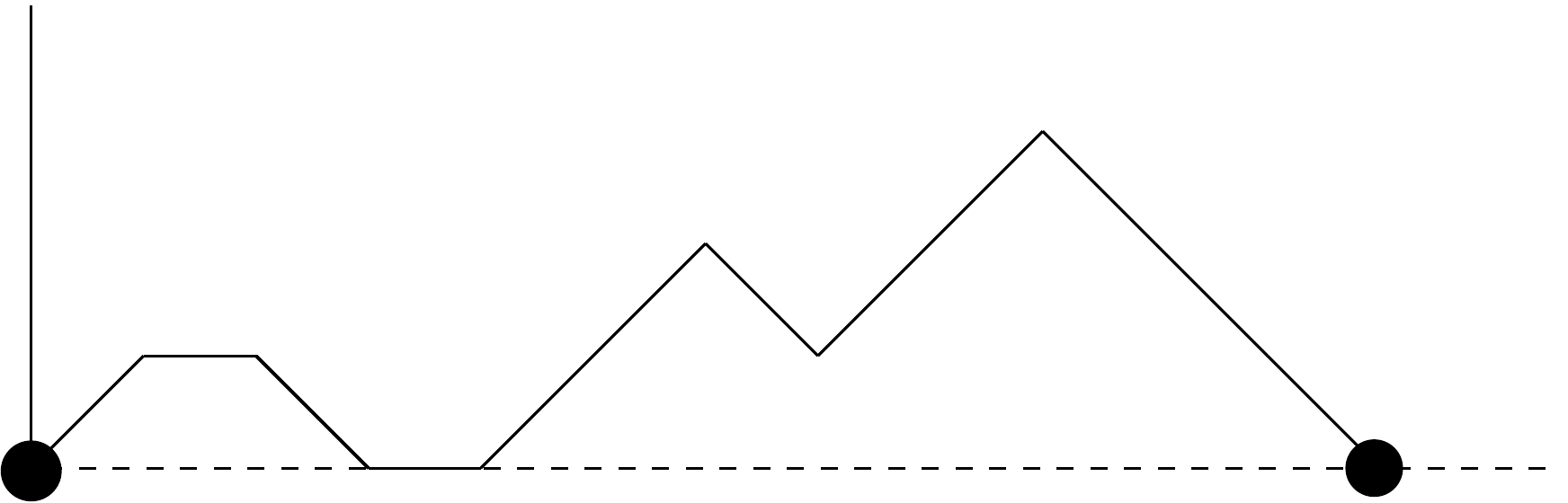}
\caption{\label{Motzkin}A Motzkin path of length $12$ which reaches a height $3$. The horizontal axis is shown dashed so the horizontal step on the axis is visible and the initial and final point are indicated}
\end{center}
\end{figure}
Since only the first
components of $\bra{W}$ and $\ket{V}$ are non-zero the paths start and end on the horizontal axis such as the path shown in Fig.\,(7). The horizontal steps may be composed into two different
``colours'', weighted by $(1+aq^n)z / (1-q)$ and $(1+bq^n)z/(1-q)$ respectively, so the model is really one of bi-coloured Motzkin paths, which 
are 
of considerable interest to combinatorialists  because may be put into bijection
with numerous other objects. 

In the lattice path model the dominant paths contributing to the normalization in the  $HD$ and $LD$ phases contain mostly one colour of
step bound closely to the horizontal axis, where the binding energies
are given in terms of $\alpha$ and $\beta$. The maximal current phase, on the other hand, corresponds to unbound paths in which entropy dominates \cite{Blythe_review}. Restricting $C$ to be finite in this context means that the  heights of the paths in the ensemble are restricted to lie below a ceiling determined by the dimension of the representation. It is therefore no surprise that finite dimensional representations, which cannot describe unbound paths, do not see the maximal current phase.  

The generating function
of $Z_N$ in the q-harmonic oscillator representation could also be viewed
as a Motzkin lattice path model since the matrices there are still tridiagonal, but in this case the form of the vectors in equ.\,(\ref{qvec}) means that the paths may start and end at {\it any} height, showing why a consistent truncation to  height zero and one paths does not appear to be possible in that case.

\section{Conclusions}

We have seen that a  two-dimensional representation of the PASEP quadratic algebra at $q=-1$ presents broadly reasonable physical behaviour, although the oscillating sign exponential corrections at both ends are unusual. We compared its properties in detail with the  two-dimensional representation which exists  when $abq=1$ and $0<q<1$.  The behaviour when the boundary conditions were extended to non-zero
$\gamma$ and $\delta$ was found to be similar, in that two-dimensional physical
representations existed when parameters were restricted appropriately,
e.g. $abq=1$ or $abcd=1$, alongside another two-dimensional representation when $q=-1$.

We  found that a naive continuation to $q=-1$ was not possible for the representation of the PASEP  related to the q-harmonic oscillator because of the form of the vectors $\bra{W}$,$\ket{V}$ in that case and presented some heuristic arguments from a weighted lattice path interpretation
of the PASEP  as to why such behaviour might have been expected.

Exploring   $n$-dimensional representations for $q^n=1, \, n>2$ would be a possible extension of the current work, but we note that in such cases the $Z_N$ is no longer necessarily real when expressed as a function of real $\alpha,\beta$ so extracting a physical interpretation for currents, densities and correlators may not be quite so straightforward. Having already allowed the unphysical value of $q=-1$ in the two-dimensional representation,
one might also take unphysical values of $a,b$ in that case. It would then be possible to arrange $\lambda_+ = \lambda_- = 1$, and hence a  phase transition, at $a=b=i$.
Another avenue which might be pursued is the
use of larger representations with $q^n=1$, $n \gg 2$ to approach the symmetric limit of the PASEP at $q=1$ through large finite dimensional representations of the PASEP algebra.  
 
A further task is, of course, finding a plausible physical setting for the two-dimensional $q=-1$ representation of the PASEP  discussed here. We have been unable to do so  ourselves thus far.
 
%%%%%%%%%%% ACKNOWLEDG(E)MENTS %%%%%%%%%%%

\section{Acknowledgements}
M. S. Stringer's work on the (partially) asymmetric exclusion process (PASEP) was (partially) supported by a  Postgraduate Students' Allowances Scheme (PSAS) grant from the Student Award Agency for Scotland (SAAS).

\vspace{1cm}

%%%%%%%%%%%%%%% REFERENCES %%%%%%%%%%%%%%% 

% hand-modified bibtex output

\end{document}